\documentclass[reprint,superscriptaddress,showpacs,twocolumn,groupedaddress,amsmath,amssymb,aps,prl]{revtex4}
\usepackage{graphicx}  % needed for figures
\usepackage{dcolumn}   % needed for some tables
\usepackage{bm}        % for math
\usepackage{tabularx}
\usepackage{array}

\usepackage{float}
\usepackage[caption = false]{subfig}

\usepackage{color} %color words
\usepackage{ulem} %\uline,\uuline,\uwave,\sout,\xout
\usepackage{hyperref}

\begin{document}

\title{Converting a topologically trivial superconductor into a topological superconductor via magnetic doping}

\author{Wei Qin}
\affiliation{International Center for Quantum Design of Functional Materials (ICQD), Hefei National Laboratory for Physical Sciences at Microscale (HFNL), and Synergetic Innovation Center of Quantum Information and Quantum Physics, University of Science and Technology of China, Hefei, Anhui, 230026, China}

\author{Di Xiao}
\affiliation{Department of Physics, Carnegie Mellon University, Pittsburgh, Pennsylvania 15213, USA}

\author{Kai Chang}
\affiliation{SKLSM, Institute of Semiconductors, Chinese Academy of Sciences, P. O. Box 912, Beijing 100083, China}

\author{Shun-Qing Shen}
\affiliation{Department of Physics, The University of Hong Kong, Pokfulam Road, Hong Kong, China}

\author{Zhenyu Zhang}
\affiliation{International Center for Quantum Design of Functional Materials (ICQD), Hefei National Laboratory for Physical Sciences at Microscale (HFNL), and Synergetic Innovation Center of Quantum Information and Quantum Physics, University of Science and Technology of China, Hefei, Anhui, 230026, China}

\date{\today}

\begin{abstract}
We present a comparative theoretical study of the effects of standard Anderson and magnetic disorders on the topological phases of two-dimensional Rashba spin-orbit coupled superconductors, with the initial state to be either topologically trivial or nontrivial. Using the self-consistent Born approximation approach, we show that the presence of Anderson disorders will drive a topological superconductor into a topologically trivial superconductor in the weak coupling limit. Even more strikingly, a topologically trivial superconductor can be driven into a topological superconductor upon diluted doping of independent magnetic disorders, which gradually narrow, close, and reopen the quasi-particle gap in a nontrivial manner. These topological phase transitions are distinctly characterized by the changes in the corresponding topological invariants. The central findings made here are also confirmed using a complementary numerical approach by solving the Bogoliubov-de Gennes equations self-consistently within a tight-binding model. The present study offers appealing new schemes for potential experimental realization of topological superconductors.
\pacs{05.30.Rt, 74.62.En, 74.78.-w, 75.30.Hx}
\end{abstract}

\maketitle
\textit{Introduction}.---Topological superconductors (TSCs) \cite{X. L. Qi} have been intensively explored recently as candidate systems for realization of Majorana fermions \cite{E. Majorana}. Due to their exotic non-Abelian braiding statistics \cite{G. Moore,D. A. Ivanove}, Majorana fermions are in turn expected to play an important role in future fault-tolerant topological quantum computation \cite{A. Kitaev,C. Nayak}. Many different schemes have been proposed to realize TSCs, including centrosymmetric odd-parity pairing copper oxide superconductors \cite{L. Fu1,S. Sasaki}, surface states of topological insulators \cite{L. Fu2,W. Qin} or two-dimensional (2D) Rashba spin-orbit coupled semiconductors \cite{J. D. Sau1} proximity coupled with an  \textit{s-}wave superconductor, and 1D ferromagnetic Shiba chain on top of a conventional superconductor with strong spin-orbit coupling \cite{S. N. Perge}. These different innovative proposals continue to stimulate active research efforts on definitive experimental realization of TSCs and observation of Majorana fermions.

In essentially all the compelling experimental demonstrations of TSCs and/or Majorana fermions reported so far \cite{S. N. Perge,K. D. Nelson,V. Mourik,L. P. Rokhinson,M. X. Wang}, the presence of certain types of disorders had to be unavoidable. Generally speaking, disorders are undesirable physical entities, because they may destroy the salient properties of a clean system. On the other hand, there have also been numerous examples showing that properly introduced disorders can lead to the emergence of various intriguing new physical phenomena in otherwise clean but ordinary host systems. One such example is the recent discovery of the topological Anderson insulating (TAI) state in HgTe quantum wells \cite{J. Li,H. Jiang}. It was shown that the presence of Anderson disorders can drive a topologically trivial insulating state into a topologically nontrivial state (or the TAI state), with a quantized conductance of $G_0=2e^2/h$ \cite{J. Li,H. Jiang,C. W. Groth,H.-M. Guo}. In view of the widespread presence of disorders in realistic systems and the recent example on TAI, it is natural to investigate the feasibilities of disorder-induced topological phase transitions in superconducting systems, especially the potential disorder-assisted realization of TSCs.

In this Letter, we carry out a comparative study of the effects of standard Anderson and magnetic disorders on the topological phases of 2D Rashba spin-orbit coupled superconductors using complementary theoretical approaches, with the initial state to be either topologically trivial or nontrivial. First, we use the self-consistent Born approximation (SCBA) to investigate the disorder renormalized density of states (DOS), and show that the presence of usual Anderson disorders will drive a topological superconductor into a topologically trivial superconductor in the weak coupling limit. More strikingly, a topologically trivial superconductor can be driven into a topological superconductor upon doping of diluted magnetic disorders, which drive an intricate closing-and-reopening process of the quasi-particle gap in a nontrivial manner. These topological phase transitions are quantitatively characterized by their corresponding topological invariants, and the central findings are also confirmed by solving the Bogoliubov-de Gennes equations self-consistently within a tight-binding model. The present study offers new insights towards potential experimental realization of TSCs.

\textit{Theoretical model}.---To start, we consider a 2D electron gas with a Rashba-type SOC and a Zeeman field $h$. Discussions on potential experimental realizations of such systems will be deferred to near the end of this paper. By introducing a proper attractive interaction between the electrons, the system is in a superconducting state, and can further be classified as a topological superconductor when the Zeeman field exceeds a critical value $h_c$ \cite{J. D. Sau1,C. W. Zhang,X.-J. Liu}. The total Hamiltonian $H$ of such a 2D system contains two parts. The first part is the kinetic energy, given as $H_0=\int d\textbf{r} \psi^{\dagger}(\textbf{r}) \mathcal{H}_0 (\textbf{r}) \psi(\textbf{r}) $, where $\psi^{\dagger}(\textbf{r})=(\psi_{\uparrow}^{\dagger}(\textbf{r}),\psi_{\downarrow}^{\dagger}(\textbf{r}))$ and $\mathcal{H}_0 (\textbf{r})=-\hbar^2\nabla^2/(2m_e)-\mu-i \lambda (\partial_y \sigma_x-\partial_x \sigma_y)-h\sigma_z$. Here, $m_e$ is the effective mass of the electrons, $\mu$ is the chemical potential, $\lambda$ is the strength of the Rashba SOC, and $\sigma_{i}$ with $i=x, y, z$ are the Pauli spin matrices. The second part is the on-site attractive interaction between the electrons, described as $H_a=U\int d\textbf{r} \psi_{\uparrow}^{\dagger}(\textbf{r}) \psi_{\downarrow}^{\dagger}(\textbf{r}) \psi_{\downarrow}(\textbf{r}) \psi_{\uparrow}(\textbf{r})$, where $U (<0)$ is the attraction strength. In the momentum space, by treating the two-body interaction $H_{a}$ within the mean-field approximation, the total Hamiltonian $H$ can be written as $H=1/2\sum_{k}\Psi^{\dagger}_k \mathcal{H}(\vec{k}) \Psi^{\dagger}_k$, with
\begin{equation}
\mathcal{H}(\vec{k})=(\epsilon_k-\mu) \tau_z+\lambda (k_y\sigma_x-k_x\sigma_y)-h\sigma_z+\Delta \sigma_z\tau_x,
\end{equation}
where $\Psi^{\dagger}(k)=(c_{k\uparrow}^{\dagger},c_{k\downarrow}^{\dagger},c_{-k\downarrow},c_{-k\uparrow})$ are the field operators in the Nambu spinor basis, $\epsilon_k=\hbar^2k^2/2m_e$, $\tau_i$ with $i=x, y, z$ are the Pauli matrices acting on the particle-hole degrees of freedom, and $\Delta=U\sum_{k} \langle c_{-k\downarrow} c_{k\uparrow}\rangle$ is the mean-field superconducting order parameter. The system described by Eq.~(1) is a topologically non-trivial superconductor when $h>h_c=\sqrt{\Delta^2+\mu^2}$ \cite{J. Alicea}. Here we differentiate the present model systems, where time-reversal symmetry is explicitly broken by the Zeeman term, with those considered in some earlier studies in which time-reversal symmetry is protected \cite{K. Michaeli,F. Zhang}.

To describe the interactions between the electrons and disorders introduced into the system, we consider a local scattering Hamiltonian $H_{im}=-\int d\textbf{r} \psi^{\dagger}(\textbf{r}) \hat{V}_{im}  \delta(\textbf{r}-\textbf{r}_0) \psi(\textbf{r})$, whose simple form is able to capture the essential physics to be exploited in the present study. In the case of the usual Anderson disorders, $\hat{V}_{im}$ is the on-site disorder potential $V$ distributed uniformly in the interval $(-V_0, V_0)$. For magnetic disorders, we choose an isotropic form given by $\hat{V}_{im}=J(\vec{S} \cdot \vec{\sigma})$, where $J$ denotes the exchange coupling strength between the electrons and magnetic disorders, $\vec{\sigma}=(\sigma_x, \sigma_y, \sigma_z)$ for the electron spin, and $\vec{S}$ is the moment of a magnetic dopant. In Nambu notations, the disorder scattering Hamiltonian can be rewritten as
\begin{equation}
H_{im}=-\frac{1}{2}\sum_{kk'} \Psi_k^{\dagger}  \hat{\mathcal{V}}_{im} \Psi_{k'},
\end{equation}
where $\hat{\mathcal{V}}_{im}=V\tau_z$ for the Anderson disorders and $\hat{\mathcal{V}}_{im}=J\vec{S} \cdot \vec{\alpha}$ for the magnetic disorders. Here the effective electron spin operator $\vec{\alpha}$ is defined as $\vec{\alpha}= 1/2[(1+\tau_z) \vec{\sigma} +(1-\tau_z) \sigma_z \vec{\sigma} \sigma_z]$. Hamiltonian (2) is a general form of the interaction between the disorders and superconducting quasi-particles, and Eqs.~(1) and (2) provide the framework for further studying the disorder effects on the topological phases of the superconducting systems.

The Green's function formalism is well suited to study the effects induced by multiple disorders in a superconductor. After averaging over the randomly distributed disorders, the Matsubara Green's function of the system described by Eqs.~(1) and (2) is given as
\begin{equation}
\hat{G}^{-1}(\vec{k},\omega_n)=i\omega_n-\mathcal{H}(\vec{k})-\hat{\Sigma},
\end{equation}
where $\omega_n=(2n+1)\pi k_B T$, $k_B$ is the Boltzmann constant, $T$ represents the temperature, and $\hat{\Sigma}$ is the self-energy. In the presence of disorders, the superconducting order parameter is determined by the self-consistent equation
\begin{equation}
\Delta=-\frac{1}{4}Uk_B T\sum_{\omega_n}\sum_{k} \text{Tr}[\hat{G}(\vec{k},\omega_n)\sigma_z \tau_x].
\end{equation}
In the SCBA approach, $\hat{\Sigma}$ is given by 
\begin{equation}
\hat{\Sigma}(\vec{k},\omega_n)=n_{im}\sum_{k} \hat{\mathcal{V}}_{im} \hat{G} (\vec{k},\omega_n)\hat{\mathcal{V}}_{im} ,
\end{equation}
where $n_{im}$ is the concentration of the disorders. Taking into account of the symmetry restriction and the matrix structure of $\hat{\Sigma}$, the self-energy effects can be manifested as disorder-induced renormalizations of $\omega_n$, $\mu$, $h$, and $\Delta$. As shown in Eq.~(5), $\hat{\Sigma}$ is independent of the momentum, leading to a renormalized form of the Green's function $\hat{G}(\vec{k},\omega_n)=[i\tilde{\omega}_n-(\epsilon_k-\tilde{\mu})\tau_z-\lambda (k_y\sigma_x-k_x\sigma_y)+\tilde{h}\sigma_z-\tilde{\Delta} \sigma_z\tau_x]^{-1}$, obtained with the replacements of $\omega_n \rightarrow \tilde{\omega}_n$, $\mu \rightarrow \tilde{\mu}$, $h \rightarrow \tilde{h}$, and $\Delta \rightarrow \tilde{\Delta}$. The solution of Eq.~(5) gives rise to a set of self-consistent equations 
\begin{equation}
\begin{aligned}
\tilde{\omega}_n&=\omega_{n}+\alpha_1F_1\left(\tilde{\omega}_n,\tilde{h},\tilde{\Delta}\right)\tilde{\omega}_n,  \\
\tilde{h}&=h-\alpha_2  F_2\left(\tilde{\omega}_n,\tilde{h},\tilde{\Delta}\right) \tilde{h}, \\
\tilde{\Delta}&=\Delta+\alpha_3 F_3\left(\tilde{\omega}_n,\tilde{h},\tilde{\Delta}\right) \tilde{\Delta},
\end{aligned}
\end{equation}
where $\alpha_{1,2,3}=\frac{1}{12}n_{im}V_0^2$ for the Anderson disorders, and $\alpha_{1}=-\alpha_{3}=\frac{1}{4}n_{im}J^2S^2$, $\alpha_{2}=\frac{1}{4}n_{im}J^2(S_x^2+S_y^2-S_z^2)$ for the magnetic disorders. Here we have derived the detailed analytical formulas for $F_{1,2,3}(\tilde{\omega}_n,\tilde{h},\tilde{\Delta})$ from Eq.~(5) by considering the matrix structure of the Green's function (given in the Supplemental Material \cite{SM}). We also ignore the renormalization of $\tilde{\mu}$ as it simply changes the chemical potential, which can be preset suitably, unlike the TAI cases, where the renormalization of $\tilde{\mu}$ has significant effects \cite{H. Jiang,C. W. Groth,H.-M. Guo}. The disorder-averaged Green's function $\hat{G}(\vec{k},\omega_n)$ can then be obtained by solving Eqs.~(6) and (4) self-consistently.

\textit{Topological phase transitions}.---To study the disorder effects, we investigate the DOS defined as $\rho(\omega)=-(1/2\pi) \text{Im}\{\text{Tr}[\hat{G}_r(\omega) \sigma_0\tau_0]\}$, where $\hat{G}_r(\omega)$ is the retarded Green's function obtained by performing analytical continuation on $\hat{G}(\omega_{n})$. According to Anderson's theorem \cite{P. W. Anderson}, a conventional \textit{s-}wave superconductor can be barely influenced by potential disorders (later known as Anderson disorders), but such an impurity has been shown more recently to be able to induce mid-gap bound states in a TSC \cite{H. Hu,J. D. Sau2}. It is therefore worthwhile to study the effects of multiple Anderson disorders in TSCs relative to that in normal superconductors.
Furthermore, a magnetic impurity can lead to the appearance of the Yu-Shiba states \cite{L. Yu,H. Shiba} inside the superconducting gap even if the superconducting state is topologically trivial, making it more appealing to study the effects of multiple magnetic impurities in different classes of superconductors.

Here we first investigate the effects of multiple Anderson disorders on the topological phases in TSCs. Based on the study of the DOS, we recover a topological phase transition from the TSC state to a trivial superconducting state as illustrated in Fig.~1(a). Qualitatively, an increase in the strength $V_0$ of the on-site Anderson disorders will gradually close the quasi-particle gap of the TSC. When $V_0$ crosses a critical value $V_c$, the quasi-particle gap reopens, and the system adopts a trivial superconducting state. These findings suggest that, as far as Anderson disorders are concerned, a cleaner superconducting system is better for potential realization of TSC. Further increment of $V_0$ to the strong disorder regime is believed to transform the system to the Anderson localized state \cite{X. M. Cai}. 
\begin{figure}
\includegraphics[width=\columnwidth]{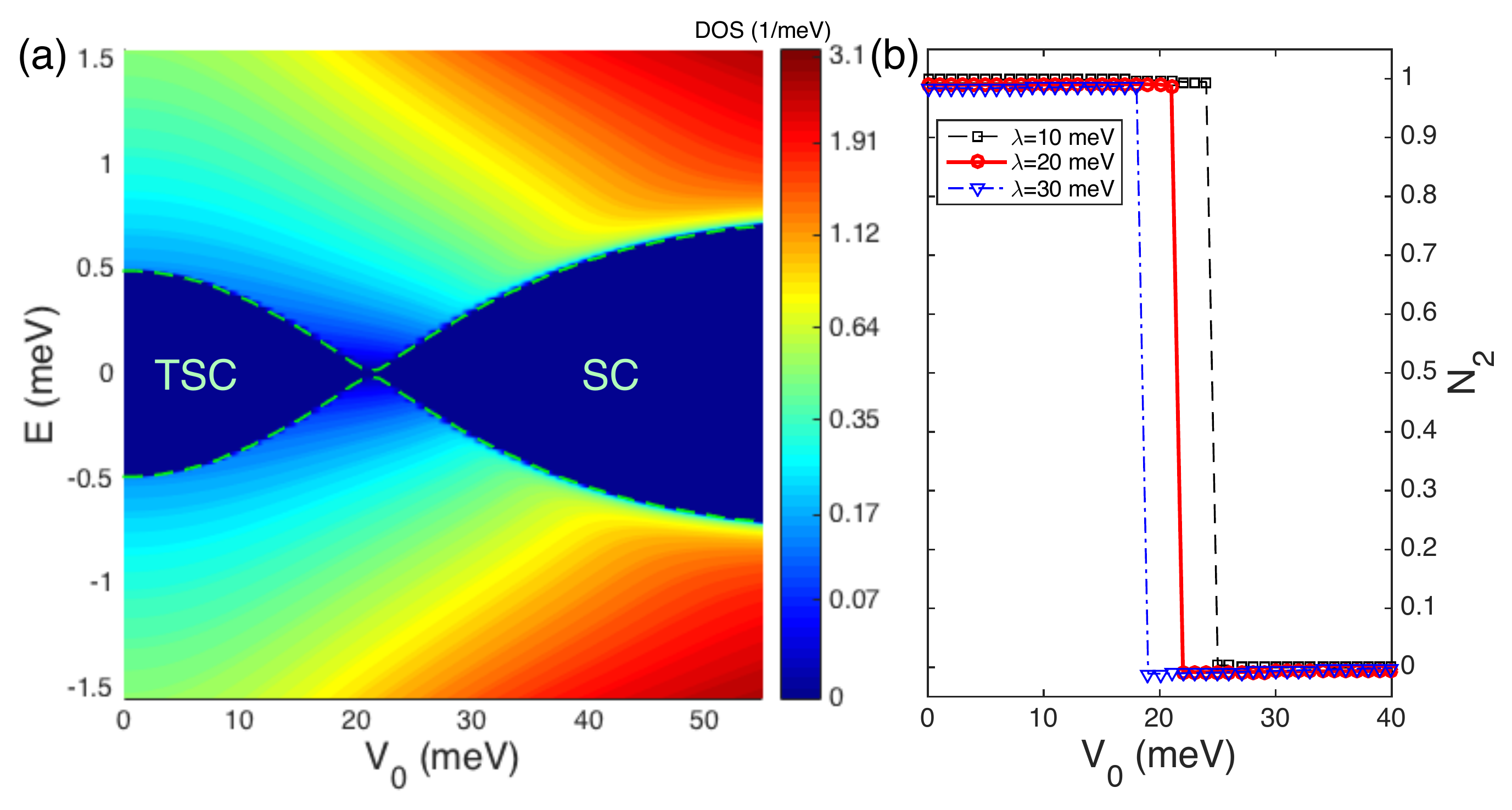}
\label{fig:local density of states}
\caption{(Color online) Topological phase transitions induced by the Anderson disorders, with the initial TSC state obtained by choosing $h=3.5$ meV. (a) The DOS as a function of the disorder strength $V_0$. The dashed (green) curves highlight the edges of the quasi-particle gap. (b) Topological invariant as a function of $V_0$ at different $\lambda$, with the solid (red) curve corresponding to the phase transition shown in (a). Other parameters include $m_e=0.01$ meV$\cdot$\AA$^{-2}$, and $U=160$ meV, leading to $\Delta\approx 3$ meV.}
\end{figure}

We next study the effects of the magnetic impurities. Unlike the case of the Anderson disorders, the presence of the magnetic disorders in a TSC cannot drive any topological phase transitions, indicating that the TSC state is relatively robust against magnetic disorders. Remarkably and most strikingly, we find that an initial topologically trivial superconductor can be converted into a TSC via diluted doping of randomly distributed magnetic impurities. As shown in Fig.~2(a), the system undergoes a gap closing and reopening process upon increasing the concentration $n_{im}$ of the magnetic disorders. A critical concentration $n_{c}$ for such a quantum phase transition is clearly identifiable. Another feature of Fig.~2(a) is that the quasi-particle gap closes again upon further increasing $n_{im}$. In this gapless region containing sufficient magnetic impurities (with $n_{im}>1.2\%$ in the present study), the system is a gapless superconductor with non-vanishing superconducting order parameter \cite{A. V. Balatsky}. However, below this concentration regime, we do have a sizable range of $n_{im}$, during which TSCs can be observed.

\begin{figure}
\includegraphics[width=\columnwidth]{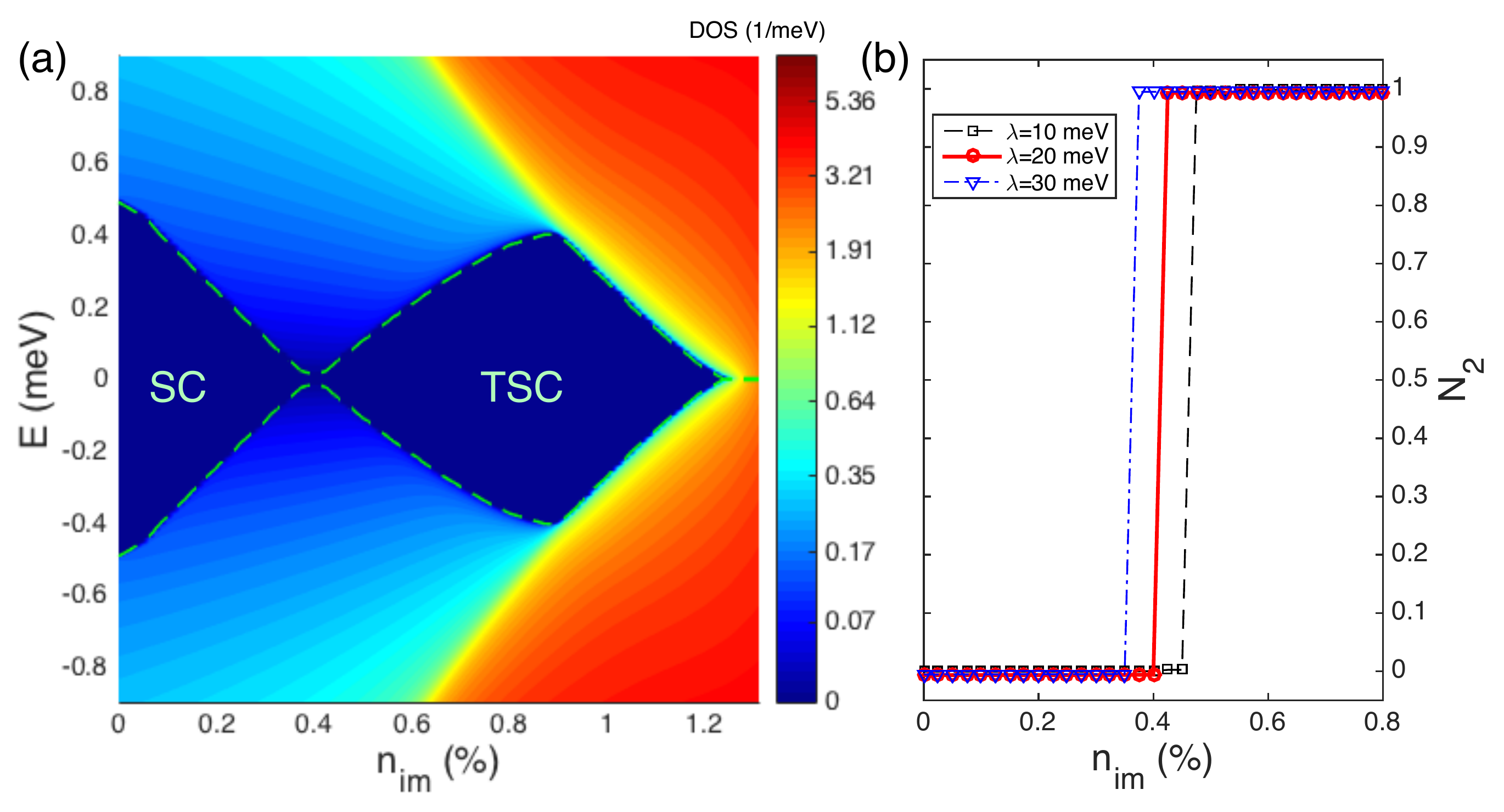}
\label{fig:local density of states}
\caption{(Color online) Topological phase transitions induced by the magnetic disorders, with the initial trivial SC state obtained by choosing $h=2.5$ meV. (a) The DOS as a function of $n_{im}$ with $J=100$ meV. The dashed (green) curves highlight the edges of the quasi-particle gap. (b) Topological invariant as a function of $n_{im}$ at different $\lambda$, with the solid (red) curve corresponding to the transition shown in (a). Other parameters include $S=\sqrt{6}$, $m_e=0.01$ meV$\cdot$\AA$^{-2}$, and $U=160$ meV, leading to $\Delta\approx 3$ meV.}
\end{figure}

To measure the occurrence of these topological phase transitions induced by disorders, we further calculate the topological invariant for various parameter regions. Due to the presence of the disorders, their interaction with the host system makes it impossible to use the standard Berry phase approach based on a pure band picture in calculating the Chern number \cite{D. Xiao}. Here we adopt an alternative and more general formula for evaluating the topological invariant, which relies on the full Green's function of the interacting system as \cite{G. E. Volovik,Z. Wang,V. Gurarie}:
\begin{equation}
\mathcal{N}_2=\frac{\pi}{3}\int \frac{d^3p}{(2\pi)^3} \text{Tr}[\epsilon_{\alpha \beta \gamma} \hat{G} \partial_{\alpha} \hat{G}^{-1}\hat{G} \partial_{\beta} \hat{G}^{-1}\hat{G} \partial_{\gamma} \hat{G}^{-1}],
\end{equation}
where $\epsilon_{\alpha \beta \gamma}$ is the Levi-Civita symbol, $p=(\omega_n,\vec{k})$, $\partial_{\alpha}=\partial_{p_{\alpha}}$, and the summations over $\alpha, \beta, \gamma$ are implied. The full Green's function of such a disordered system can be obtained self-consistently, thus the topological invariant is a straightforward calculation using Eq.~(7). As shown in Fig.~1(b), the topological invariant jumps from $\mathcal{N}_2=1$ to 0 at the critical value of $V_0$, which indicates a topological phase transition from a chiral TSC to a trivial SC. Figure~2(b) further confirms that a topologically trivial SC ($\mathcal{N}_2=0$) can be converted to a chiral TSC ($\mathcal{N}_2=1$) via increasing magnetic doping. In addition, we also find that the enhancement of the SOC strength will promote these topological phase transitions in both cases of the Anderson and magnetic disorders. Collectively, these results offer strong evidence for the existence of rich topological states induced by proper choices of disorders. 

\textit{Numerical BdG solutions}.---In this section, we numerically investigate the disorder-induced effects within a corresponding tight-binding model of the system described by Eq.~(1). By projecting Hamiltonian (1) on a 2D square lattice, we reach
\begin{equation}
\begin{aligned}
H=&-t\sum_{\langle ij\rangle \sigma }(c_{i\sigma}^{\dagger}c_{j\sigma}+\text{H.c.})+\lambda \sum_{\langle ij\rangle} (e^{i\theta_{ij}}c_{i\uparrow}^{\dagger}c_{j\downarrow}+\text{H.c.})\\
&-h\sum_{i\sigma \sigma'} (\sigma_z)_{\sigma \sigma'} c_{i\sigma}^{\dagger}c_{i\sigma'}-U\sum_{i}\hat{n}_{i\uparrow}\hat{n}_{i\downarrow}-\mu \sum_{i\sigma}\hat{n}_{i\sigma}\\
&+\sum_{i\sigma \sigma'} (\hat{V}_{im})_{\sigma \sigma'} c_{i\sigma}^{\dagger}c_{i\sigma'},
\end{aligned}
\end{equation}
where $t$ is the nearest-neighbor hopping term, $c_{i\sigma}^{\dagger}$ and $c_{i\sigma}$ are the creation and destruction operators for an electron with spin $\sigma$ on site $\textbf{r}_i$ of the square lattice, $\hat{n}_{i\sigma}=c_{i\sigma}^{\dagger}c_{i\sigma}$ is the particle number operator, and $\theta_{ij}$ is the angle between $(\textbf{r}_j-\textbf{r}_i)$ and the $x$ axis. Here, we solve the disordered system characterized by Eq.~(8) within the standard mean-field BdG approach, $\mathcal{H}_{\text{BdG}} \Psi_n(\textbf{r}_i) =E_n \Psi_n (\textbf{r}_i)$, where 
\begin{equation}
\mathcal{H}_{\text{BdG}}=
\begin{pmatrix}
\mathcal{H}^{\uparrow}_{s}(\textbf{r}_i)&\mathcal{R}&0 & \Delta(\textbf{r}_i) \\
\mathcal{R}^{\dagger}& \mathcal{H}^{\downarrow}_{s}(\textbf{r}_i) &-\Delta(\textbf{r}_i)& 0 \\
0 & -\Delta^*(\textbf{r}_i)& -\mathcal{H}^{\uparrow}_{s}(\textbf{r}_i)& \mathcal{R}^{\dagger}\\
\Delta^*(\textbf{r}_i) & 0  & \mathcal{R} & -\mathcal{H}^{\downarrow}_{s}(\textbf{r}_i)
 \end{pmatrix} 
\end{equation}
is the BdG Hamiltonian, $E_n$ is the eigenenergy of the corresponding quasi-particle wave function $\Psi_n(\textbf{r}_i)=[u_n^{\uparrow}(\textbf{r}_i),u_n^{\downarrow}(\textbf{r}_i), \nu_n^{\uparrow}(\textbf{r}_i),\nu_n^{\downarrow}(\textbf{r}_i)]^T$ defined in the Nambu spinor space. Here $\mathcal{H}_s^{\sigma}(\textbf{r}_i) u_n^{\sigma}(\textbf{r}_i)=-t\sum_{\hat{\delta}} u_{n}^{\sigma}(\textbf{r}_i+\hat{\delta})+\eta (V_{im}-\tilde{\mu}_{i\sigma}) u_{n}^{\sigma}(\textbf{r}_i)$, $\mathcal{R}(\textbf{r}_i) u_n^{\sigma}(\textbf{r}_i)=\lambda \sum_{\hat{\delta}} e^{i\theta}u_n^{-\sigma}(\textbf{r}_i+\hat{\delta})-V_{im}^{+}u_{n}^{\sigma}(\textbf{r}_i)$, $\hat{\delta}=\pm \hat{\textbf{x}},\pm \hat{\textbf{y}}$, $\Delta(\textbf{r}_i)$ is the local superconductor order parameter, and $\tilde{\mu}_{i\sigma}=\mu+U\langle \hat{n}_{i,-\sigma}\rangle$ is the modified chemical potential due to Hartree shift. We note that $V_{im}^{+}=J(S_x+iS_y)$ for the magnetic disorders and vanishes for the Anderson disorders. Starting with some initial guess values of $\Delta(\textbf{r}_i)$, we first numerically solve the BdG Hamiltonian on a $N \times N$ square lattice with periodic boundary conditions. Next, we calculate the local pairing amplitudes and particle density given by
\begin{equation}
\Delta(\textbf{r}_i)=U\sum_n u_n^{\downarrow}(\textbf{r}_i) \nu_n^{\uparrow*}(\textbf{r}_i),~~~~n_{i\sigma}=\sum_{n} |\nu_{n}^{\sigma}(\textbf{r}_i)|^2,
\end{equation}
and iterate this process until self-consistent values of $n_{i\sigma}$ and $\Delta(\textbf{r}_i)$ at each site are achieved.

\begin{figure}
\includegraphics[width=\columnwidth]{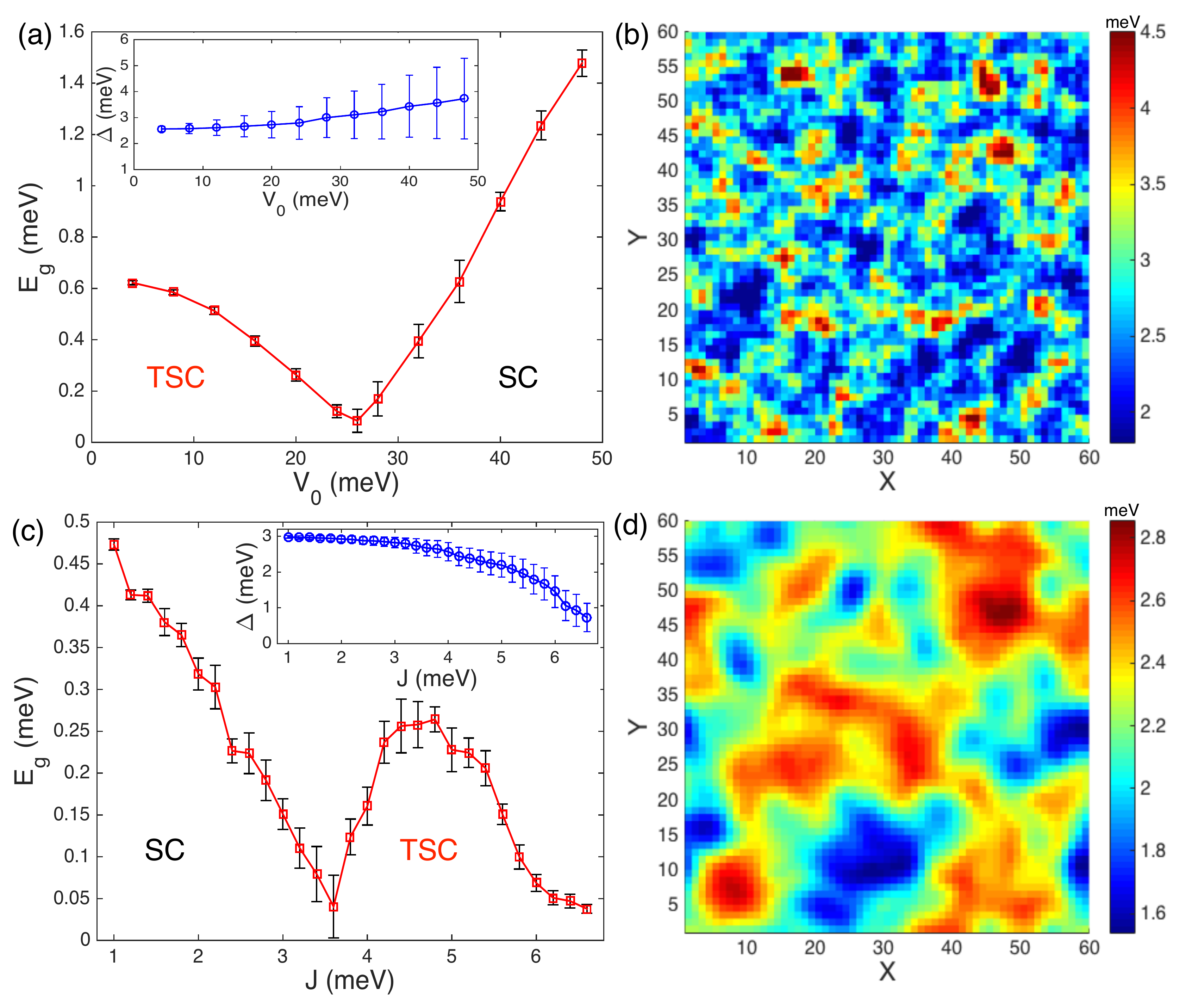}
\label{fig:numerical}
\caption{(Color online) The upper and lower panels show the numerical results of a $60\times 60$ square lattice in the presence of the on-site Anderson and magnetic disorders, respectively. (a) Quasi-particle gap as a function of the disorder strength $V_0$. (b) The spatial variations of $\Delta(\textbf{r}_i)$ for the Anderson disorders with $V_0=10$ meV. (c) Quasi-particle gap as a function of the on-site magnetic coupling strength $J$.  (d) The spatial variations of $\Delta(\textbf{r}_i)$ for randomly distributed magnetic disorders with $J=4.8$ meV. The error bars in (a) and (c) show the standard deviations of the mean for 10 samples, and the inserts give the averaged local superconducting order parameters. Other parameters include $S=\sqrt{6}$, $U=160$ meV, and $t=50$ meV, corresponding to $m_e=0.01$ meV$\cdot$\AA$^{-2}$.}
\end{figure}

Numerical solutions of the BdG equation can yield the whole quasi-particle spectrum in the presence of the disorders. As shown in Figs.~3 (a) and (c), there exists two regimes of the disorder strength showing nonzero quasi-particle gaps. Similar to the SCBA results, we also observe gap closing and reopening processes by increasing the strength of the disorders. These behaviors are qualitatively the same as that shown in Fig.~1(a) and Fig.~2(a). From the quantitative perspective, the critical value of $V_0$ for the Anderson disorders is about 26 meV, which is comparable to $V_c=22$ meV shown in Fig.~1(a). For computational convenience, the magnetic disorders are treated as on-site spins with randomly oriented directions in the numerical calculations. As shown in Fig.~2(a), the critical concentration characterizing the topological phase transition is $n_{c}=0.4\%$, translating into $J=3.2$ meV for on-site disorders, and the value is comparable to the critical coupling strength $J_c=3.6$ meV illustrated in Fig.~2(c). Based on these qualitative and quantitative comparisons between the results obtained analytically earlier and numerically now, we can conclude affirmatively that rich topological states can indeed be induced by properly introducing disorders. In particular, a topologically trivial SC can be readily converted into a TSC upon diluted doping of independent magnetic impurities. In addition, the spatial variations of $\Delta(\textbf{r}_i)$ shown in Figs.~3(b) and (d) demonstrate that the systems evolve into phase-separated regions in the presence of disorders, and the sizes of these regions are comparable to the superconducting coherence length $\xi=\nu_F/\Delta \sim 10 a$, where $\nu_F$ is the Fermi velocity, and $a$ is the lattice constant.

Before closing, we briefly discuss candidate systems for potential experimental realizations of such disorder-induced topological phase transitions. We could consider a superconductor thin film with strong SOC, such as Pb \cite{S. N. Perge,S. Y. Qin,T. Zhang} or PbBi alloyed films grown on semiconducting substrates. Due to the lacking of the inversion symmetry, the Rashba-type SOC will be present on the surface of such superconductors. Therefore, by directly doping magnetic elements into or on the surface of such 2D superconductors, their topologically trivial nature may be converted into TSCs. Our results may also be observed in the recently realized SOC coupled 2D ultracold atomic Fermi gases of $^{40}$K \cite{P. Wang} and $^6$Li atoms \cite{L. W. Cheuk}, which are shown to be topological superfluids. In addition, the present findings can be extended to 1D and 3D cases.

In summary, we have studied, using complementary analytical and numerical approaches, the effects of the Anderson and magnetic disorders on the topological phases of 2D superconductors with Rashba SOC. We have found that the presence of the Anderson disorders will drive a TSC into a topologically trivial SC in the weak coupling limit. More strikingly, a topologically trivial SC can be converted into a TSC upon diluted doping of independent magnetic impurities, which is characterized by an intricate nontrivial gap closing and reopening process. The central findings can offer new insights towards potential experimental realization of TSCs.

This work was supported by the National Natural Science Foundation of China (Grants Nos. 11034006,  61434002, and 11434010), the National Key Basic Research Program of China (Grant No. 2014CB921103), US National Science Foundation (Grant No. EFRI-1433496), and the Research Grants Council of Hong Kong (Grant No. HKU703713P).

\end{document}